# Structural consequences of the coupling of gravitons

Walter Gessner
Unterer Katzenbergweg 7
D-97084 Wuerzburg

## Abstract

The hypothetical but in string theories predicted gravitons and their allied quanta of the linearized gravitation field (G) yield, when coupled with other particles (T), a remarkable structure of the entire state space S and an autonomous concept of observables:

It is shown that S is spanned by a family of continuously many, pairwise orthogonal, pre-Hilbert spaces, called sectors. Each sector is the tensor product of an eigenspace of the operator of the 4-momentum distribution within the state space of (T), and a corresponding eigenspace within the state space of (G), both with any spatial distribution of the total 4-momentum as the eigenvalue 4-vector. Thus the states from any sector and so far the entirety of 4-momentum densities appear to be inseparably dressed by the quanta of (G). Furthermore, the special structure of the sectors leads to an own concept of observables, all of which commute from the beginning with the operator of the 4-momentum distribution.

# 1. Introduction

(T) is supposed as a quantum field theory in the Schrödinger picture with the pre-Hilbert state space $S_T$ assumed as the Fock representation [01] of the canonical commutation relations of the dynamical field variables of (T). It is further assumed that (after averaging by test functions) an operator $T_{\mu\tau}(\underline{x})$ of the energy-momentum tensor and a Hamiltonian $H_T$ do exist.

On the other hand, the canonical quantization of the linearized gravitation field [02] of the general theory of relativity leads to a theory (G) of gravitons [03] and of other quanta of the gravitation field. The Fock representation of the corresponding commutation relations gives a state space $S_G$ with an indefinite metric which must be repaired by a certain Lorentz-Hilbert operator condition LHC generalizing the classical Lorentz gauge of QED and the Lorentz-Hilbert gauge of the classical linearized gravitation field.

$T_{\mu\tau}(\underline{x})$ allows the coupling of (T) and (G) to a theory (T, G). The coupling constant is the $\kappa$ from the field equations of general relativity. The question is now to what extent the particular structure of $S_G$, especially induced by LHC, determines the structure of the state space $S_{TG}$ of (T, G). Accordingly, the physical meaning of this modified structure is to be discussed. At last, it will be checked which operators are in particular adapted to this structure so that one may call them observables.

# 2. The dynamical variables

The linearization of the field equations of general relativity in the flat Minkowski space leads to the wave equations $\Box \gamma_{\mu\tau}(x) = \kappa T_{\mu\nu}(x)$ for the symmetric tensor field $\gamma_{\mu\tau}(x)$ which describes the small deviations of the real metric tensor from the constant $g_{\mu\nu}$ normed by $\text{diag}(g_{\mu\nu}) := (-1,+1,+1,+1)$. To quantize $\gamma_{\mu\tau}(x)$ in the Schroedinger picture, time-independent, unrenorma-lized and with respect to the indices symmetric formal operators $\gamma_{\mu\tau}(\underline{x})$ are introduced by [04 – 07]

$$\gamma_{\mu\tau}(\underline{x}) := \int d^3k\, D(\underline{k})[e^{ikx} a_{\mu\tau}(\underline{k}) + e^{-ikx} a_{\mu\tau}(\underline{k})^+]. \tag{01}$$

$D(\underline{k}) := (16\pi^3|\underline{k}|)^{-1/2}$ collects the usual relativistic weight factor and the $\pi$-factors from the Fourier transformation. The creation- and annihilation operators in (01) satisfy the commutation relations [04 - 06]

$$[a_{\mu\tau}(\underline{k}), a_{\delta\varepsilon}(\underline{l})^+] = [g_{\mu\delta}g_{\tau\varepsilon} + g_{\mu\varepsilon}g_{\tau\delta} - g_{\mu\tau}g_{\delta\varepsilon}]\delta(\underline{k} - \underline{l}), \tag{02}$$

$$[a_{\mu\tau}(\underline{k}), a_{\delta\varepsilon}(\underline{l})] = 0 = [a_{\mu\tau}(\underline{k})^+, a_{\delta\varepsilon}(\underline{l})^+]. \tag{03}$$

Instead of the $a_{\mu\tau}(\underline{k})$, $a_{\mu\tau}(\underline{k})^+$, similar operators may be introduced which are closer to physical interpretations: One gets then creation and annihilation operators of gravitons and of other quanta of the linearized gravitation field, here called "Newton quanta", because they are closely related to the density of energy and momentum as will be seen. Analogously one has in QED creation and annihilation operators of photons and of scalar resp. longitudinal "Coulomb quanta" [09] related to charge densities.

The commutation relations (02, 03) lead to a Fock space $S_G$ with a Lorentz invariant but indefinite metric. Especially to overcome this unphysical feature of $S_G$, the condition LHC is introduced.

The dynamical variables of (T) satisfy canonical commutation relations. The corresponding state space $S_T$ is the Fock representation of these relations and assumed as a pre-Hilbert space. In $S_T$ creation and annihilation operators of the quanta of (T) may be defined and $T_{\mu\tau}(\underline{x})$ could be written by them. But it is only supposed that $T_{\mu\tau}(\underline{x})$ is well defined in $S_T$ (after averaging by test functions) and symmetric. Formally it is introduced

$$\Theta_{\mu\tau}(\underline{k}) := D(\underline{k}) \int d^3x\, e^{-ikx} T_{\mu\tau}(\underline{x}). \tag{04}$$

A special role play the $\Theta_{0\mu}(\underline{k})$, especially in the form

$$P_\mu(\underline{k}) := \kappa D(\underline{k})\, \Theta_{0\mu}(\underline{k}) = \kappa D(\underline{k})^2 \int d^3x\, e^{-ikx} T_{0\mu}(\underline{x}), \tag{05}$$

the Fourier transforms of the 4-momentum densities.

The Hamiltonian of (T, G) can be splitted up as [06 - 08]

$$H = H_T + H_G + H_{TG}. \tag{06}$$

$H_T$ is the Hamiltonian of (T), $H_G$ the Hamiltonian of the free gravitons and Newton quanta:

$$H_G = \tfrac{1}{2} \int d^3k |\underline{k}| [a_{\mu\tau}(\underline{k})^+ a^{\mu\tau}(\underline{k}) - \tfrac{1}{2} a(\underline{k})^+ a(\underline{k})]. \tag{07}$$

$H_{TG}$ describes the interaction between (T) and (G) and is given by

$$H_{TG} := \kappa \tfrac{1}{2} \int d^3k [\Theta^{\mu\tau}(\underline{k}) a_{\mu\tau}(\underline{k})^+ + \Theta^{\mu\tau}(\underline{k})^+ a_{\mu\tau}(\underline{k})] -$$

$$- \kappa \tfrac{1}{4} \int d^3k [\Theta(\underline{k}) a(\underline{k})^+ + \Theta(\underline{k})^+ a(\underline{k})], \tag{08}$$

where $a(k) := g^{\mu\tau} a_{\mu\tau}(\underline{k})$ and $\Theta(\underline{k}) := g^{\mu\tau} \Theta_{\mu\tau}(\underline{k})$. $\qquad(09)$

## 3. Interaction postulate and LHC

LHC uses the term [04 – 07]

$$\gamma_{\mu\tau}(\underline{x})^* := \int d^3k\, D(\underline{k}) e^{i\underline{k}\underline{x}} a_{\mu\tau}(\underline{k}) \tag{10}$$

which consists of annihilation operators only. The formal derivation $\partial^r \gamma_{r\tau}(\underline{x})^*$, $r = 1,2,3$, and $\partial^0 \gamma_{0\tau}(\underline{x})^* = i[\gamma_{0\tau}(\underline{x})^*, H]$ yields

$$\partial^\mu \gamma_{\mu\tau}(\underline{x})^* = i \int d^3k\, D(\underline{k})\, e^{i\underline{k}\underline{x}} [L_\tau(\underline{k}) - P_\tau(\underline{k})], \text{ where} \tag{11}$$

$L_\tau(\underline{k}) := D(\underline{k}) k^\mu a_{\mu\tau}(\underline{k})$, and

$$P_\tau(\underline{k}) := \kappa D(\underline{k})\, \Theta_{0\tau}(\underline{k}) = \kappa D(\underline{k})^2 \int d^3x\, e^{-i\underline{k}\underline{x}}\, T_{0\tau}(\underline{x}). \tag{12}$$

5At first, $L_\tau(\underline{k})$ and $P_\tau(\underline{k})$ are defined only formally. To get operators it is necessary to introduce test funktions $f = f(\underline{x})$ from the Schwartz space $S(\mathbb{R}^3)$ for example, so that

$$\partial \gamma_\tau[f] := \int d^3x\, f(\underline{x})\, \partial^\mu \gamma_{\mu\tau}(\underline{x})^* = L_\tau[f] - P_\tau[f] \text{ with} \qquad (13)$$

$$L_\tau[f] := \int d^3k\, f(\underline{k})\, L_\tau(\underline{k}), \text{ and } P_\tau[f] := \int d^3k\, f(\underline{k})\, P_\tau(\underline{k}). \qquad (14)$$

$P_\tau[f]$ is the operator of a distribution of the total 4-momentum of (T), whereas $L_\tau[f]$ acts in the space of gravitons and Newton quanta $S_G$. The set $P[f]^*$ collects all the 4-vector expectation values $<a|P_\tau[f]|a>$ with $a \in S_T$ and $\|a\| = 1$.

The following postulate combines the usual interaction postulate and LHC:

## Postulate:

$S_{TG}$ is spanned by all maximal pre-Hilbert spaces $s$ satisfying

i) $s = k \otimes l$ where $k$ and $l$ are pre-Hilbert subspaces of $S_T$ resp. $S_G$. (15)

ii) $\partial \gamma_\tau[f] (k \otimes l) = 0$ for all $\tau$ and all $f \in S(\mathbb{R}^3)$. (16)

To study the consequences of this postulate one needs the following lemma:

## Lemma

Let be given a pre-Hilbert space $S$ and a symmetric operator $\Omega: S \to S$. If $<\varphi|\Omega|\varphi> = 0$ for all $\varphi \in S$, then $\Omega = 0$.

## Proof:

Assume that $\varphi, \psi \in S$ do exist so that $<\varphi|\Omega|\psi> \neq 0$. Then $<\varphi|\Omega|\psi>$ can be written as $<\varphi|\Omega|\psi> = e^{i\alpha}|<\varphi|\Omega|\psi>|$ with $\alpha \in [0, 2\pi[$.

The element $\varphi + e^{i\beta}\psi \in S$, where $\beta$ is any parameter, satisfies now

$$0 = <\varphi + e^{i\beta}\psi|\Omega|\varphi + e^{i\beta}\psi> = e^{-i\beta}<\psi|\Omega|\varphi> + e^{+i\beta}<\varphi|\Omega|\psi> = \qquad (17)$$
$$e^{+i(\alpha+\beta)}|<\varphi|\Omega|\psi>| + e^{-i(\alpha+\beta)}|<\varphi|\Omega|\psi>| = 2\cos(\alpha+\beta)|<\varphi|\Omega|\psi>|$$

so that $\cos(\alpha+\beta) = 0$.

The choice $\beta := -\alpha$ for example leads to a contradiction. Therefore

$$<\varphi|\Omega|\psi> = 0 \text{ for all } \varphi, \psi \in S. \qquad (18)$$

If now an element $\psi \in S$ exists with $\Omega\psi \neq 0$ then $0 \neq \langle\Omega\psi|\Omega\psi\rangle = \langle\varphi|\Omega\psi\rangle = \langle\varphi|\Omega|\psi\rangle$ with $\varphi := \Omega\psi \in S$. Contradiction. ∎

**Theorem:**

As a consequence of the postulate, $S_{TG}$ contains to any $f \in S(\mathbb{R}^3)$ continuously many pairwise orthogonal pre-Hilbert spaces $s^{(p)}$, called „sectors", with

$$s^{(p)} = k^{(p)} \otimes l^{(p)}. \tag{19}$$

$k^{(p)}$ is the eigenspace (0-space included) of $P_\tau[f]$ in $S_T$, $l^{(p)}$ the eigenspace of $L_\tau[f]$ in $S_G$, both with the same eigenvalue 4-vector $(p) := p_\tau[f] \in P[f]^*$.

**Proof:**

Let $\varphi := a \otimes \alpha \in s$ with $a \in k$, $\alpha \in l$. Then $\partial \gamma_\tau[f](a \otimes \alpha) = 0$ reads $(P_\tau[f]a) \otimes \alpha = a \otimes (L_\tau[f]\alpha)$ for all $\tau$. The multiplication with $a \otimes \alpha$ yields $\langle a|P_\tau[f]|a\rangle\langle\alpha|\alpha\rangle = \langle\alpha|L_\tau[f]|\alpha\rangle\langle a|a\rangle$. With $\|a\| = \|\alpha\| = 1$ this reads

$$\langle a|P_\tau[f]|a\rangle = \langle\alpha|L_\tau[f]|\alpha\rangle, \tag{20}$$

so that $\langle a|P_\tau[f]|a\rangle$ does not depend on $a \in k$, and $\langle\alpha|L_\tau[f]|\alpha\rangle$ does not depend on $\alpha \in l$. Let now be $p_\tau[f] := \langle a|P_\tau[f]|a\rangle \in P[f]^*$ and

$$\Omega_\tau[f] := P_\tau[f] - p_\tau[f]. \tag{21}$$

Then $\langle a|\Omega_\tau[f]|a\rangle = 0$ for all $a$, and, because of the lemma, $\Omega_\tau[f]a = 0$ for all $a \in k$. This means $P_\tau[f]a = p_\tau[f]a$ and thus $a \in k^{(p)}$.

The situation in $S_G$ is more complicated because $S_G$ has an indefinite metric as a consequence of the commutation relations (02), (03). But in [04 – 06] it is shown that $S_G$ contains a lot of positive-semidefinite subspaces L, especially the eigenspaces of $L_\tau[f]$. Any such L leads then to the pre-Hilbert space L/N where $N := \{\alpha \in L| \langle\alpha|\alpha\rangle = 0\}$. The quotient space L/N can be embedded in L so that $S_G$ contains pre-Hilbert spaces as assumed in the postulate.

The same conclusion as above leads then to the eigenspaces $l^{(p)} \cong L^{(p)}/N^{(p)}$ with

$$L^{(p)} := \{\alpha \in S_G | L_\tau[f]\alpha = p_\tau[f]\alpha \text{ for all } \tau\} \text{ and} \tag{22}$$

$$N^{(p)} := \{\alpha \in L^{(p)}| \langle\alpha|\alpha\rangle = 0\}. \tag{23}$$

∎

Theorem 1 provides to any f continuously many pairwise orthogonal sectors $s^{(p)}$. Until now this conglomerate as a whole has no pre-Hilbert space structure. It seems that LHC and the interaction postulate are not strong enough to force such a structure. Therefore, additional conditions are to be introduced: If, for example, the theory can be restricted to a countable set of (p), $S_{TG}$ is the direct sum of the $s^{(p)}$. Another possibility consists in the introduction of a $\sigma$ – finite measure $\mu$. The $\varphi \in S_{TG}$ can then be written as

$$\varphi = \int \sqrt{d\mu}\, \varphi^{(p)} \text{ with } \varphi^{(p)} \in s^{(p)} \text{ and } <\varphi|\varphi> = \int d\mu <\varphi^{(p)}|\varphi^{(p)}>.$$

Summarizing, the postulate does not lead to an obvious definition of $S_{TG}$ as a pre-Hilbert space. Rather it provides only the sectors and their pairwise orthogonality as a general basis for the construction of $S_{TG}$. Although the postulate only sets general conditions, all in this way possible state spaces have some common characteristics:

- Each element $\varphi \in S_{TG}$ consists of components $\varphi^{(p)} \in s^{(p)}$, where (p) stands in the closest relationship to a 4-momentum density. Thus $\varphi$ appears as the superposition of sharp momentum densities and not as a formal term only.

- To any density of the 4-momentum, generated by the particles of (T), the space $S_G$ has in store a suitable configuration ("robe") of Newton quanta. $S_{TG}$ couples this configuration inseparably with the 4-momentum density (dressing of the 4-momenta). Probably this dressing of the 4-momenta contributes to the "dark matter".

- On $S_{TG}$ both operators $P_\tau[f]$ and $L_\tau[f]$ are identical. This can be regarded as a **dualism** with practical consequences: The difficult (T)-dependent $P_\tau[f]$ can be replaced in each case by the formally uncomplicated $L_\tau[f]$.

## 4. Observables

Even if the metric of $S_{TG}$ is still open, observables must be compatible with the sector structure of $S_{TG}$. Therefore any observable $\Omega: S_{TG} \to S_{TG}$ has to be symmetric (resp. selfadjoint) and to satisfy

$$[\Omega, \partial \Gamma_\tau[f]] = 0 \text{ for all } \tau \text{ and all } f \in S(\mathbb{R}^3). \tag{24}$$

Examples: 4-momentum and angular momentum:
In the $\underline{x}$-space such a large finite area A is selected that all processes of a closed physical system within a fixed time interval held in it. By a test function $f_A(\underline{x})$ with $f_A(\underline{x}) = 1$ for all $\underline{x} \in A$ the total 4-momentum of the system can than be written as

$$P_\tau := \int d^3x \, f_A(\underline{x}) \, T_{0\tau}(\underline{x}). \tag{25}$$

Instead of $T_{0\tau}(\underline{x})$ the operator $P_\tau(\underline{k})$ may be introduced and replaced by $L_\tau(\underline{k})$ because of the sector structure of $S_{TG}$:

$$P_\tau := \kappa' \int d^3k |\underline{k}| f_A(\underline{k}) \, L_\tau(\underline{k}) \tag{26}$$

with a constant $\kappa'$. In this way, the 4-momentum allows two completely different observables. While (25) depends crucially on the T-quanta, the uncomplicated term (26) is independent of them because it operates in the space $S_G$ of gravitons and Newton quanta.

By $\underline{P}(\underline{x}) := (T_{0r}(\underline{x}))$, $r = 1,2,3$, of the density of the 4-momentum, the total angular momentum of the system is given by ($\underline{0} \in A$)

$$\underline{J} := \int d^3x \, f_A(\underline{x})[\underline{x} \times \underline{P}(\underline{x})]. \tag{27}$$

$\underline{J}$ is observable in the above sense and has a dual form also, operating in the space $S_G$ of gravitons and Newton quanta.
The equations
$$\partial \gamma_\tau[f_A] := L_\tau[f_A] - P_\tau[f_A] \text{ and } [\Omega, \partial \Gamma_\tau[f_A]] = 0 \tag{28}$$
lead in a natural way to the consequences, that any observable $\Omega$ satisfies ($\tau = 0,1,2,3$)

$$[\Omega, P_\tau] = [\Omega, \underline{J}] = 0. \tag{29}$$

This is a well known theorem proven in [10] with respect to a different but more general concept of observables [11].

## References

[01] S. S. Schweber, An Introduction to Relativistic Quantum Field Theory (Harper & Row, New York, 1964)

[02] Ch. Misner, K. S. Thorne, J. A. Wheeler, Gravitation (Freeman, San Franzisco, 1973)

[03] C. Rovelli, arXiv: gr-qc/0006061 (2001)

[04] S. N. Gupta, Proc. Phys. Soc. (London) A $\underline{65}$, 161, 608 (1952)

[05] J. Gomatam, Phys. Rev. D $\underline{6}$, 1292 (1971)

[06] H. P. Dürr and E. Rudolph, Nuovo Cimento A $\underline{62}$, 411 (1969)

[07] D. Witt: The Quantization of Geometry, in L. Witten: Gravitation, an Introduction to current Research, New York 1962

[08] A. Ashtekar and R. Geroch, Rep. Prog. Phys. 37, 1211 (1974)

[09] W. Gessner and V. Ernst, J. Math. Phys. $\underline{21}$, 93 (1980)

[10] F. Strocchi and A. S. Wightman, J. Math. Phys. $\underline{15}$, 2198 (1974)

[11] R. Haag and D. Kastler, J. Math. Phys. $\underline{5}$, 848 (1964)